\documentclass{pasa}%
\title[Polarization and collisions]{Scattering polarization of the $d$-states of  ions and solar magnetic field: 
Effects of isotropic collisions}
\author[Derouich et al.]{M. Derouich$^{1,2}$\thanks{Corresponding author: M. Derouich (Astronomy Department, Faculty of Sciences, King Abdulaziz University, 21589 Jeddah, Saudi Arabia)}, H. Basurah$^1$, B. Badruddin$^1$ \\
\affil{$^1$Astronomy Department, Faculty of Sciences, King Abdulaziz University, 21589 Jeddah, Saudi Arabia}%
\affil{$^2$Sousse University,  ESSTHS, Lamine Abbassi street, 4011 H. Sousse, Tunisia}}%
\jid{PASA}
\doi{10.1017/pas.\the\year.xxx}
\jyear{\the\year}

\usepackage[authoryear]{natbib}
\bibpunct{(}{)}{;}{a}{}{,}
\usepackage{aas_macros}
\usepackage{hyperref} 
\hypersetup{colorlinks,citecolor=blue,linkcolor=blue,urlcolor=blue}

\begin{document}
\begin{abstract}
Analysis of solar magnetic fields using  observations as well as theoretical interpretations   of the scattering polarization  is commonly designated as a high priority area of the solar research.  The interpretation of the observed polarization   raises  a serious theoretical challenge to the researchers involved in this field. In fact,  realistic interpretations  need detailed investigations of the depolarizing role of   isotropic collisions with neutral hydrogen.

The goal of this paper is to determine new relationships  which   allow the calculation of any   collisional rates  of the $d$-levels of ions   by simply determining the value of $n^*$ and $E_p$ without the need  of determining the interaction potentials and treating the dynamics of collisions. The determination of $n^*$ and $E_p$ is easy and based on atomic data usually available online. Accurate collisional rates allow a reliable diagnostics of solar magnetic fields.

In this work  we applied our collisional FORTRAN code to a large number of cases involving complex and simple ions. After that, the results are utilized and injected in a genetic programming code developed with C-langugae in order to infer original relationships which will be of great help to   solar applications.
We   discussed the accurarcy of our collisional rates in the  cases  of  polarized  complex atoms and atoms with hyperfine structure.  The relationships are expressed on the   tensorial basis and we explain how to include their  contributions in the master equation giving the variation of the density matrix elements.  

As a test, we compared the results obtained through the general  relationships provided in this work with the results obtained 
directly by running our code of collisions. These comparisons show a percentage of error of about 10\% in the average value. 
Our results could be   implemented easily  in numerical codes concerned with the  simulations of the scattering polarization to obtain accurately the magnetic field in the quiet Sun.  
\end{abstract}
\begin{keywords}
polarization  --  collisions -- line: formation  --  Sun: atmosphere -- Sun: magnetic fields
\end{keywords}
\maketitle%
\section{INTRODUCTION }
\label{sec:intro}

  The  second solar spectrum (SSS) is the spectrum of the scattering linear polarization observed close to the limb of the quiet Sun where the anisotropy of the radiation   is maximum. The SSS is depolarized by turbulent solar magnetic fields via the Hanle effect and by isotropic collisions with neutral hydrogen.  The Hanle depolarization  allows   diagnostics of the magnetic fields by confronting  the discrepancy between the polarization modeled in the absence of    magnetic fields and the  observed   polarization (e. g. Stenflo 1982;  Landi Degl'Innocenti  1983; Sahal-Br\'echot et al 1986; Stenflo 2004; Trujillo Bueno et al. 2004; Derouich et al. 2006; Faurobert et al. 2009).

The study of the  widths, shifts and (de)polarization of spectral lines under the influence of collisions with neutral perturbers has already been the subject of numerous studies (e.g. Lemaire et al. 1985; Baird et al. 1979; Monteiro et al. 1988; Chambaud \& L\'evy 1989; Anstee \& O'Mara 1991; Spielfiedel et al. 1991; Krsljanin \& Peach 1993; Leininger et al. 2000; Derouich et al. 2003a; Derouich et al. 2015, see also Barklem 2016 and references therein). 
In addition,  Anstee, Barklem and O'Mara (ABO)   developed a powerful semi-classical theory during the 1990s for
collisional line broadening by neutral hydrogen (Anstee \& O'Mara 1991, 1995; Anstee 1992; 
 Anstee et al. 1997; Barklem 1998; Barklem \& O'Mara 1997; Barklem et al. 1998).  The ABO theory has been   generalized successfully by Derouich, Sahal-Br\'echot and Barklem (DSB)  to obtain   the depolarization and polarization transfer rates by collisions with neutral hydrogen  (see  for example  Derouich et al.  2003a and Derouich 2004).  

In this context, we developed an original collisional code which enables us to provide a large amount of the collisional depolarization and  polarization transfer  rates (e.g. Derouich et al. 2015, Derouich et al. 2003a).
As done in most  semi-classical collisional approaches, the approximation of the rectilinear trajectories is adopted in the collisional method of this work. This approximation was validated by
several works (see, for example, Smith et al., 1969; Allard \& Kielkopf 1982).  Furthermore, we assume that  the impact approximation   is valid where we consider only  the binary collisions which are well separated and uncorrelated. This approximation is  well satisfied in the solar conditions (see, e.g., Derouich et al. 2003a and Derouich 2004).  

The DSB and ABO theories allowed us to obtain  widespread collisional data for  {\it neutral} atoms (Derouich et al. 2015). Therefore,  depolarization and polarization transfer rates can be determined from general relationships without the requirement to utilize the collisional numerical code     level by level. Unfortunately, this is not possible to do  for   {\it  ionized} atoms where one must proceed   level by level. Thus,   any calculation  of the depolarization
and transfer of polarization rates is limted to a given level of a given ionized atom.  

 Recently, Derouich (2017) removed this limitation   in the case of $p$-levels of ions. In this paper,  we intend to extend the results of Derouich (2017) to the case of $d$-states.

The paper is organized as follows. Section 2 states the problem to be treated. In Section 3, we describe the numerical work and provide the general relationships for the $d$-states of simple ions.    Atoms with hyperfine structure and complex atoms  are presented in sections   4 and 5. Finally, the conclusions of the paper are presented in section 6.

\section{Statement of the problem}
For neutral atoms, the   Uns\"old energy $E_p$ has a constant value  E$_p$ = -4/9 a.u.   (see, e.g., Anstee 1992, Barklem \& O'Mara 1998; Derouich et al.  2003a). The constant   E$_p$ = -4/9 a.u. is used in the expression of the interaction potential which, in turn, is used to calculate  the collisional rates for any level on any atom.  On the other hand, for ionized atoms, the Uns\"old energy is not constant   and one has to determine the value of   E$_p$ for each case (Derouich et al.  2004). After that the code of collisions should be run  to get the collisional rates level by level.   This is a strong restriction which hinders  the generalization of the DSB results   published in Derouich et al. (2004).  That restriction has been  removed for the $p$-states of ions in the recent paper of  Derouich  (2017). 
 The model presented in   Derouich  (2017) is extended here  to d $(l = 2)$  ionic levels. 

We can summarize the calculations of the collisional depolarization rates for the levels of ionized atoms in two main stages. In the first  stage  one calculates  the value of $E_p$   for the ionic state under study. This calculation requires summation over  values of   the dipole oscillator strengths of all transitions to the level of interest   for the perturbed ion and to the ground state   for the neutral hydrogen atom  (see Derouich et al. 2004 and references therein).  Then, in the  second stage,   the calculated value of E$_p$ is injected in the code of collisions to determine the interaction potential.  Once the interaction potential is obtained, it is  introduced in the formalism treating the dynamics of collisions  in order to calculate the probabilities of collisions on the tensorial basis   (see equation 10 of Derouich et al. 2003b).  Depolarization rates are obtained after    integrating  the probabilities of collisions, firstly over the impact-parameter $b$ and secondly over a Maxwell distribution of velocities $f(v)$, for a local temperature $T$ of the solar atmosphere. 

In practice,   solar physicists  can obtain  the  value of  $E_p$ (stage 1) but it is   difficult  for them to determine the interaction potential and treat the dynamics of collisions in order to obtain   the collisional depolarization rates    (stage 2). Our problem can be stated as how to use powerful numerical methods in order to overcome   the difficulty  raised in the stage 2. 
The essential contribution  of this  paper is to   allow  solar community to   compute easily the collisional rates for the $d$-states of any ion by completing only stage 1.   
\section{General relationships for $d$-states of simple ions}
\subsection{Definitions and notations}
Polarization  of atomic   levels quantifies the population differences  and the interferences   between the magnetic sublevels.
The values of the polarization of    $J$-levels are typically  presented   by the  values of the elements $^{J}\rho_{q}^{k}$  of the density matrix -- the index $k$ gives the tensorial order inside the level where $0 \le k \le 2J$ and $q$ concerns the infeterferences between 
the sublevels, $-k \le q \le +k$. 

The tensorial order and thus the polarization  of atomic   levels  can be generated  by anisotropic illumination. Consequently, one receives polarized light   emitted by ions containing these levels. To obtain theoretically the Stokes parameters characterizing the radiation   ($I$, $Q$, $U$, and $V$), one needs to solve the radiative transfer problem coupled to the statistical equilibrium equations (SEE) for the  elements  $^{J}\rho_{q}^{k}$   (see, e.g.,   Landi Degl'Innocenti \&  Landolfi 2004).     The  elements  $^{J}\rho_{q}^{k}$ at each point of the LOS are the unknowns of the problem.
 To obtain these elements precisely one must  solve the SSE and take into account all relevant     processes intervening in the time of the  polarized line formation.  These processes are mainly the collisional inetractions, 
magnetic fields effects and radiative transfer effects. 

As it is showed in Equation (1) of Derouich et al. (2004),  the variation of     $ \rho_{q}^{k} (J)$     due to isotropic collisions is:  
\begin{eqnarray} \label{eq_LL}
\big[\frac{d \; \rho_q^{k} (\ J)}{dt}\big]_{coll} & = &  
  - \big[\sum_{J' \ne J} \zeta (J \to  J') + D^k(J) \big] \times \rho_q^{k} (J) \nonumber \\
&& + \sum_{J' \ne J} 
C^k(J' \to  J)   \times \rho_q^{k} (J')   
\end{eqnarray}
where  $\zeta (J \to  J')$ are the fine structure transfer rates  given by (Equation 5 of Derouich et al. 2003b):
\begin{eqnarray} \label{eq_1}
\zeta (J \to  J')  & = &   \sqrt{ \frac{2J'+1} {2J+1}} \times
C^0(J \to  J').    
\end{eqnarray}

In the absence of the polarization,  $k=0$ and  $q=0$ and using the general density matrix formalism one shows that
the unknowns of the problem become  $\rho_0^{0} (\alpha J)$  which are simply proportional to the populations of the levels 
as in spectroscopic studies  (e.g. Asplund 2005).
  
 In the solar   photosphere  where the   SSS is originated,   the isotropic collisions with neutral hydrogen  dominate any other kind of collisions. These collisions occur inside the same electronic ionic level.   The depolarization and polarization transfer rates are $q$-independent because the collisions are isotropic 
 (e.g. Sahal-Br\'echot 1977; Derouich et al. 2003a).  
 
 We denote    the  depolarization   rates  due to purely elastic collisions occuring in only one $J$-level by $D^{k}(J)$; the expression of $D^{k}$ is given  by equations (7) and (9) of Derouich et al. (2003a).   The $C^k(J \to J')$    indicate the polarization transfer rates due to  collisions between the initial  level  $(J)$ and the final level $(J')$  (see Equation 3 of Derouich et al. 2003b). 
 

 Our results are concentrated on  the  $d$-states with   orbital momentum $l$=2. The spin of the optical electron is  $s$=1/2.  The total angular momentum  is  {\bf $J$}={\bf $l$}+${\bf s}$   and thus one has $J$=3/2 or $J$=5/2. As a result, there are two fine structure  states   $^2D_{J=\frac{3}{2}}$ and   $^2D_{J=\frac{5}{2}}$.

Throughout this paper, all the collisional rates are given in s$^{-1}$. In the framework of   the impact approximation, the rates are proportional to the neutral hydrogen density in cm$^{-3}$ denoted by $n_{\textrm {\scriptsize H}}$ and they depend on the  temperature $T$  given in Kelvins. We showed in our previous works that  the collisional  
rates  can be expressed as (e.g. Derouich et al. 2003b;  Derouich et al. 2004)
\begin{eqnarray} \label{eq_1}
D^k(J,T) & = &  D^k(J,T=5000K) \times    (\frac{T} {5000})^{ \frac{1-\lambda} {2}} \\
C^k(J \to J',T) & = &  C^k(J \to J',T=5000K) \times    (\frac{T} {5000})^{ \frac{1-\lambda} {2}}  \nonumber \\
\end{eqnarray}
where $T=5000$K  and $\lambda$ is the so-called velocity exponent. Both DSB and 
 ABO found  that  collisional rates have typical   temperature dependence of $T^{0.38}$ which means that  $\lambda \eqsim  0.25$. 
 
 \subsection{Numerical results}
 The numerical code of collisions is updated in order to make it  useful for calculations of large number of collisional rates of ionic $d$-states.
Using this code,  collisional rates are obtained at $T=5000K$ for   effective principal quantum 
number $n^*$ in the interval [2.5, 4].  Furthermore, we computed for each value of  $n^*$,  grids of collisional rates for each Uns\"old energy $E_p$   which is in the typical interval [-2, -0.6] (in atomic units).  We adopt a step size of 0.1 in the variation of  $n^*$ and $E_p$. Consequently, we obtain   three dimensional tables containing the   collisional rates  with the parameters $n^*$ and 
$E_p$.

Based on these tables, we used our Genetic Programming (GP)  method of fitting in order to provide analytically general relationships between  collisional rates, 
 $n^*$ and $E_p$. This method has been used successfully in Derouich et al. (2015) and in  Derouich   (2017). 
 The GP method minimizes the summed square of   the difference between the tabulated rates  and rates obtained from the  GP  relationships.  The relationships  predicted by the GP-based model are compared with the available  data of the $d$-states of the Ca II, Ba II and Sr II  in order to estimate their precision. 
  
  The following equations show the relationships between the collisional depolarization/transfer of polarization rates of the left-hand side and the $n^*$ and $E_p$ in the right-hand side.  Let us recall that, for  $J=\frac{3}{2}$,   $0 \le k \le 3$ and for  $J=\frac{5}{2}$,   $0 \le k \le 5$. In  particular, by definition,   $D^{k=0}(J)$=0. In the following, we provide   the non-zero depolarization and polarization transfer rates associated to the levels  $^2D_{J=\frac{3}{2}}$ and $^2D_{j=\frac{5}{2}}$.
   \subsubsection{Depolarization rates of the level  $^2D_{J=\frac{3}{2}}$}
 \begin{itemize}
\item
\begin{eqnarray}  \label{eq_1}
D^{1}(\frac{3}{2}) (T=5000K)/(n_{\textrm {\scriptsize H}}  \times 10^{-9}) =  \nonumber \\  \frac{3./Y-(1.+Y)}{(X+1.)/(X-2.)+(X-2.)} \nonumber \\   +  \frac{2. \times X}{(8.-X)} \times (2. \times X+X/3.-3.)  
\end{eqnarray}

 where  $X=n^{*}_{p}$    and $Y=  -E_p >0$. 
 
 In order to compare with published results, let us consider the case of the $d$-state of the Ca II where $n^{*}_{d}=2.315 $ and $E_{p}$=-1.2.   According to the Equation 14 of Derouich et al. (2004), the reference value of $D^{1}(\frac{1}{2}) (T=5000K)$/$(n_{\textrm {\scriptsize H}}  \times 10^{-9})$=1.9904  s$^{-1}$=$D_\textrm{reference}$. The GP methods provide Equation (\ref{eq_1}) which implies that  $D^{1}(\frac{1}{2}) (T=5000K)$/$(n_{\textrm {\scriptsize H}}  \times 10^{-9})$=1.98365   s$^{-1}$=$D_\textrm{GP}$. Thus the   relative error  is
  \begin{eqnarray}  \label{eq_11}
re\textrm{(Ca II)}= \frac{|D_\textrm{reference}-D_\textrm{GP}|}{D_\textrm{reference}} \times 100=0.5 \%
\end{eqnarray}
  which demonstrates how good is the  precision of the GP method of fitting.   
\item
 \begin{eqnarray}  \label{eq_11}
D^{2}(\frac{3}{2}) (T=5000K)  /(n_{\textrm {\scriptsize H}}  \times 10^{-9})  = \nonumber \\  X^2-3. +  \frac{5./Y-X}{(3.-Y)+5./X}  \\   - \frac{2.}{[(3.  \times  X-3.)   +(Y-3.)/X]} \nonumber
\end{eqnarray}

With   respect to the reference value given in Equation 14 of Derouich et al. (2004), $re\textrm{(Ca II)}$ $\simeq$ 10 \%, in the case of the $d$-state of the Ca II.  In addition, in the case of the $d$-state of the Sr II, $re\textrm{(Sr II)}$ $\simeq$ 23 \%. For the Ba II of  the $d$-state, $re\textrm{(Ba II)}$ $\simeq$ 7\%. To determine the error bars for the Sr II and Ba II, we compared  with the results obtained by Derouich (2008) and Deb \& Derouich (2014).
\item
 \begin{eqnarray}  \label{eq_11}
D^{3}(\frac{3}{2}) (T=5000K) /(n_{\textrm {\scriptsize H}}  \times 10^{-9})  && =  \nonumber   \\   \frac{X^2}{(2.  \times  X+2.  \times  Y)}+[X-(1.+Y)]  \times  [\frac{(X-2.)}{14.}] &&
\nonumber \\    +X-2.  &&
\end{eqnarray}
By comparing with the reference value of $D^{3}(\frac{3}{2}) (T=5000K)$ given in Equation 14 of Derouich et al. (2004), we get $re\textrm{(Ca II)}$   \% $\simeq$ 0.25 \%.
   \subsubsection{Depolarization rates of the level  $^2D_{J=\frac{5}{2}}$}
\item
 \begin{eqnarray}  \label{eq_D15demi}
&& D^{1}(\frac{5}{2}) (T=5000K) /(n_{\textrm {\scriptsize H}}  \times 10^{-9})   =   \nonumber   \\ 
& & [6./(1.+Y)+6.-X] \times (X-2.) \\ 
& & +\frac{2.\times X-4.-2./(1.+X)}{5. \times (3. \times X/2.-7./2.)}  \nonumber 
\end{eqnarray}
At $T=5000$K, $re\textrm{(Ca II)}$    $\simeq$  11 \% for  the level $^2D_{5/2}$ of the Ca II.  
\item
 \begin{eqnarray}  \label{eq_D25demi}
 D^{2}(\frac{5}{2}) (T=5000K) /(n_{\textrm {\scriptsize H}}  \times 10^{-9})  = \nonumber \\   (4./5. -7./X+3.  \times    X)-  \\   
  \frac{3.  \times    X}{X^2-3.+X}-\frac{(5.-X/Y)}{(7./3.+2./Y)} \nonumber
\end{eqnarray}
We find that, $re\textrm{(Ca II)}$    $\simeq$ 20 \%  for Ca II, $re\textrm{(Ba II)}$   $\simeq$  12 \%  for Ba II, and $re\textrm{(Sr II)}$    $\simeq$ 1.3 \%  for Sr II.
\item
 \begin{eqnarray}  \label{eq_D35demi}
D^{3}(\frac{5}{2}) (T=5000K) /(n_{\textrm {\scriptsize H}}  \times 10^{-9})   =   \nonumber \\   11./80.  \times    X^2  \times    (X+2.)   \\    - \frac{Y}{(X-Y/7.-7./(X  \times    4.))} \nonumber 
\end{eqnarray}
where   $re\textrm{(Ca II)}$  is    less than  0.1 \%. 
\item
 \begin{eqnarray}  \label{eq_11}
D^{4}(\frac{5}{2}) (T=5000K) /(n_{\textrm {\scriptsize H}}  \times 10^{-9})  =   \nonumber \\   1./Y+(X-2.)  \times    (6.-X/3.)- \\  
  \frac{\frac{5.}{(5.  \times    X-Y  \times    X)  \times    (Y/5.+5.)}}{\frac{(Y-7.)}{5.}+2.  \times    X^2-4.  \times    X} \nonumber 
\end{eqnarray}
with $re\textrm{(Ca II)}$   $\simeq$    5 \%. 
\item
 \begin{eqnarray}  \label{eq_11}
D^{5}(\frac{5}{2}) (T=5000K) /(n_{\textrm {\scriptsize H}}  \times 10^{-9})  =   \nonumber \\    -\frac{18}{5.}-\frac{Y}{X} + \frac{13.}{7.}  \times  X +\frac{X^2}{5.}
\end{eqnarray}
and $re\textrm{(Ca II)}$   is less than    3 \%. 
\end{itemize}
   \subsubsection{Polarization transfer rates between the  levels  $^2D_{J=\frac{3}{2}}$ and $^2D_{J=\frac{5}{2}}$}
Concerning the non-zero polarization transfer rates  between the levels   $^2D_{J=\frac{3}{2}}$ and $^2D_{J=\frac{5}{2}}$, only the rates $C^{0}(\frac{3}{2} \to \frac{5}{2}) $, $C^{1}(\frac{3}{2} \to \frac{5}{2}) $,  $C^{2}(\frac{3}{2} \to \frac{5}{2})$, and  $C^{3}(\frac{3}{2} \to \frac{5}{2})$ are   non-zero,
\begin{itemize}
\item
 \begin{eqnarray}  \label{eq_11}
C^{0}(\frac{3}{2} \to \frac{5}{2})  (T=5000K) /(n_{\textrm {\scriptsize H}}  \times 10^{-9})  =   \nonumber \\  (X^2-\frac{83.}{20.})  \times    \frac{3.}{X}+(1.-X)  \times  Y^2  \times    \frac{X}{49.}+  \nonumber \\     \frac{5. \times (X^2-Y-3.)      }{(X-Y)  \times    X}  \times    \frac{(Y+2.)}{(X  \times    (5.+X)  \times    Y^2)} \nonumber  \\
\end{eqnarray}
By comparing to the reference value of $C^{0}(\frac{3}{2} \to \frac{5}{2})  (T=5000K))$ given in Equation 17 of Derouich et al. (2004), we found that  $re\textrm{(Ca II)}$ $\simeq$ 2 \%, $re\textrm{(Ba II)}$ $\simeq$ 19 \%, $re\textrm{(Sr II)}$ $\simeq$ 4 \%. Note that in Derouich et al. (2004), the polarization transfer rates are denoted by $D^{k}(J' \to J,T)$ instead of the notation  $C^k(J' \to J,T)$ adopted here.
\item
 \begin{eqnarray}  \label{eq_11}
C^{1}(\frac{3}{2} \to \frac{5}{2})  (T=5000K)  /(n_{\textrm {\scriptsize H}}  \times 10^{-9}) =    \nonumber   \\  \frac{7.}{5.Y(X+2.) }  \times (X-2.) +2.   X-\frac{7.}{X}
\end{eqnarray}
We notice that $re\textrm{(Ca II)}$ $\simeq$ 1 \%.
\item
 \begin{eqnarray}  \label{eq_11}
C^{2}(\frac{3}{2} \to \frac{5}{2})  (T=5000K)  /(n_{\textrm {\scriptsize H}}  \times 10^{-9}) = \nonumber   \\  2. \times X-\frac{Y}{3.}-3.-  \frac{(X-3.)}{(7.-Y)}  \nonumber \\   - \frac{\frac{7.}{X^2}}{(-2.+X)  \times (\frac{Y}{2.}+X) \times (Y+2.)}
\end{eqnarray}
We notice that $re\textrm{(Ca II)}$ $\simeq$ 5 \%, $re\textrm{(Ba II)}$ $\simeq$ 3 \% and $re\textrm{(Sr II)}$ $\simeq$ 26 \%.
\item
 \begin{eqnarray}  \label{C3d3demi-d5demi}
C^{3}(\frac{3}{2} \to \frac{5}{2})  (T=5000K)  /(n_{\textrm {\scriptsize H}}  \times 10^{-9})  = \nonumber   \\   \frac{ \frac{8. \times X}{7.}-2. }{(\frac{Y}{3.}+\frac{1.}{2.}) \times [(X+Y) \times \frac{X}{5.}+ \frac{\frac{7.}{2.}}{(X+Y)}]}   \\    +X-2 \nonumber
\end{eqnarray}
When we compare a direct calculation of $C^{3}(\frac{3}{2} \to \frac{5}{2}) $ and its calculation via Equation (\ref{C3d3demi-d5demi}), we find that $re\textrm{(Ca II)}$ $\simeq$ 7 \%.
\end{itemize}
Only the excitation  collisional transfer rates $C^{0}(\frac{3}{2} \to \frac{5}{2})$,    $C^{1}(\frac{3}{2} \to \frac{5}{2})$,  $C^{2}(\frac{3}{2} \to \frac{5}{2}) $ and $C^{3}(\frac{3}{2} \to \frac{5}{2})$ are given here. However, it is 
straightforward to retrieve the values of the   deexcitation  collisional rates 
 $C^{0}(\frac{5}{2} \to \frac{3}{2})$,    $C^{1}(\frac{5}{2} \to \frac{3}{2})$,  $C^{2}(\frac{5}{2} \to \frac{3}{2}) $ and $C^{3}(\frac{5}{2} \to \frac{3}{2})$ by applying the detailed balance relation:
 \begin{eqnarray}  \label{eq_11}
C^k(J_u \to J_l, T)  =     \\ \frac{2J_l+1}{2J_u+1}  \exp \left(\frac{E_{J_u}-E_{J_l}}{k_BT}\right) \; \; C^k_I(J_l \to J_u, T)  \nonumber
\end{eqnarray}
where $J_l$=3/2 (lower  level) and     $J_u$=5/2   (upper  level); $E_{J}$ being the energy of the level ($J $) and $k_B$ the Boltzmann constant.

Thanks to the relationships given here, any   collisional rates can be obtained by simply determining the value of $n^*$ and $E_p$.

\section{Hyperfine structure collisional rates}
 The procedure for calculating the hyperfine structure  rates consists of following steps: 
 \begin{itemize}
   \item  Determination of the hyperfine eigenstates after diagonalization of the Hamiltonian  
  \item   Calculation of the interaction potential which is expressed in  the   basis of  the  set of the  hyperfine eigenstates
    \item  Calculation of the  the collisional scattering matrix in the same hyperfine structure basis by solving the 
 Schr\"odinger equation
    \item  Finally, one must perform   the integrations over the impact parameters 
and the relative velocities. 
 \end{itemize}
 This is what we will call ``direct  method". There is also 
 an ``indirect" and more practical method that allows the calculation of the hyperfine depolarization and polarization transfer 
 rates via the values of the collisional rates associated to the fine structure,  $D^{k}(J)$ and $C^k(J \to J')$. 
 
\subsection{On the possibility of the direct method of the calculations of the hyperfine structure  rates}
There are two types of problems in the direct calculations of the hyperfine collisinal rates, first one corresponds to the zero-magnetic field case and second one corresponds to non-zero magnetic field case. 
\begin{enumerate}
  \item 
 Zero magnetic field case :  Giving a  real numerical prediction in the cases where the coherences (interferences) between different hyperfine levels  are taken into account is a   complicated problem which would be taken up in the future.     In fact, 
one needs a full study based on accurate interaction potentials and close coupling dynamical description that fully accounts for the
 collision  channels associated to the hyperfine levels and the coherences between them. In particular, the interaction potential, the wavefunctions and the Hamiltonian must be expressed in a basis vectors    formed from the complete set of hyperfine eigenstates  which take   into account the coherences.  The fact that the magnetic field is zero implies that the basis is uniquely determined and the   good quantum numbers of the basis remain unchanged\footnote{We recall that $J$ is said to be a  good  or
 conserved quantum number if every eigenvector  remains an eigenvector with the same eigenvalue, as time evolves. The total angular momentum $J$ is not a good quantum number of the total Hamiltonian  if the magnetic field $B \ne 0$.  The total  Hamiltonian should be diagonalized  for each magnetic field to yield the energies and the
eigenfunctions of the collision channels.} which makes the    calculations of the hyperfine structure  rates possible but rather complicated.  
  \item 
 Non-zero magnetic field case :   In a magnetised medium like the solar atmosphere and especially in the so-called   Paschen-Back effect regime,   the hyperfine eigenstates of the total Hamiltonian can be changed according to the value of the magnetic field. Thus, total  Hamiltonian should be diagonalized  for each magnetic field to yield  the
hyperfine eigenstates (e.g. Landi Degl'Innocenti  \& Landolfi 2004) and, after that, the interaction potential and  the dynamics of collisions have to be treated in the   basis of  the  set of the obtained hyperfine eigenstates\footnote{Interactions
with magnetic fields could perturb the internal hyperfine structure of the colliding particles,
induce couplings between states otherwise uncoupled, break the  
symmetry of the problem  and
couple the motion of the center of mass with the relative collision dynamics (e.g.     Beigman   \& Lebedev 1995; Krems  \&   Dalgarno 2004; Bivona et al. 2005).}. 

 Inversion of the scattering polarization  including a  proper   treatment of elastic and inelastic collisions in the presence of strong magnetic field cannot be  performed at least in the near future. The main reason is that the   magnetic field is unknown before the interpretation  of the scattering polarization and thus cannot be considered in the collisional problem. In fact,
the magnetic field    is typically  invoked   as   free fitting  parameter  to superpose  theoretical  Stokes  profiles to observed ones,     however  the  collisional rates  have to be obtained and introduced in the SEE before the  fitting. 

 A possible solution of this problem is to perform  the calculations of the collisional rates  for  expected range of magnetic field values   in order to  determine an empirical   law of the variation of the collisional cross section  as a function of the magnetic field (e.g. Volpi \& Bohn  2002).

The solar physicist should be aware that using collisional rates calculated in zero magnetic field are suitable only for the Hanle effect regime. It should be sufficient to observe that magnetic-induced crossing interferences in the $^2P_{\frac{3}{2}}$ level of Na I would set on   for $\sim$ 10 Gauss implying the change of the eigenstates of the 
total Hamiltonian. It is worth noting that Kerkeni \& Bommier (2002) unawarely used collisional rates for the $^2P_{\frac{3}{2}}$ level of Na I  which are calculated in the limit of zero magnetic field. This would induce  errors in their modeling of the atomic polarization of the $^2P_{\frac{3}{2}}$ level    considering that the strength of magnetic fields in the quiet Sun can be tenths of Gauss.    
 \end{enumerate}

\subsection{Calculation  of the hyperfine structure  rates by using the indirect method: the   frozen nuclear spin  approximation }
It is worth  noticing  that  the indirect method can be used only for zero magnetic field case. \\
 The (de)polarization   rates  obtained in this paper (i.e. $D^{k}(J)$ and $C^k(J \to J')$) can be utulized only in zero magnetic field case or in Hanle effect regime.
 In addition,  we do not obtain directly the collisional rates for the hyperfine levels.  We neglect the effect of the hyperfine structure during the collision. 
 
 In the DSB approach, our aim was to propose a general method, i.e. which can be applied to any ion. Thus,
we tried to neglect, whenever  possible, the parts of the  interaction potential  which are specific to a given ion.   We compared our results to those obtained 
with quantum chemistry methods and showed that   the 
precision of the DSB method is sufficient for correctly studying the second solar spectrum  (see Derouich et al. 2003a, Derouich et al. 2004).  We notice that those quantum-chemistry methods   take into account spin-orbit   interactions  but not the hyperfine structure  effects. 

Our numerical results   are obtained for the depolarization and polarization transfer rates for fine structure $J$-levels.  In order to obtain 
the collisional rates for the hyperfine levels, the   frozen nuclear spin  approximation must be applied. 
 The frozen   approximation means that   during the collision  
the  nuclear spin of the perturbed ion is conserved. 
 This  would be the case if 
\begin{eqnarray} \label{eq_HFS}
\tau \; \Delta E_{HFS} \ll 1,
\end{eqnarray}
 $\Delta E_{HFS}$ is the energy of the hyperfine structure  and $ \displaystyle \tau $ is the typical time duration of a collision. Equation    (\ref{eq_HFS}) is typically satisfied in the solar conditions (e.g. Derouich et al. 2003a).

  In the typical solar conditions  it is easy to show that $1/\tau$ $\sim$ 10$^{13}$ s$^{-1}$ (i.e. $1/\tau \sim$ 334 cm$^{-1}$).  In these conditions,  the hyperfine  splitting is usually much smaller than $1/\tau$ and therefore one can assume that the nuclear spin is conserved during the collision.  It is important, however, to  not confuse this condition with the fact that the  statistical equilibrium equations must be
solved for the hyperfine levels when the inverse of the lifetime
of the level is smaller than the hyperfine splitting, i.e. the
hyperfine levels are separated.

 Our relationships  given by the Equations (5--17) can be used to easily derive the hyperfine depolarization rates and polarization transfer rates for any singly ionized atom.  This is the indirect method which is based on the frozen core approximation. According to this method, the hyperfine collisional rates are given as a linear combination of the fine structure rates, first obtained by   Nienhuis  (1976) and   Omont (1977).


\section{Complex atoms}
\subsection{Why treating complex atoms?}
 The   important aspect of the DSB method is its extension to the cases of complex atoms and ions. Originally this extension was presented and explained in detail by Derouich et al. (2005a).   Its importance 
 appeared especially in the atlas of the linearly-polarized solar-limb spectrum published by Gandorfer  
(2000, 2002, 2005), where the linear polarization of the lines of complex atoms/ions  is particularly interesting. In fact among atomic polarized lines presented in Gandorfer (2000),   there are  38 $\%$
of polarized Fe {\sc i} lines and 13 $\%$ for Ti {\sc i}.  The observations of the linear polarization  reported by  Gandorfer (2000, 2002, 2005) show  polarization peaks in many spectral lines of  complex ions, e.g. Nd   {\sc ii}  5249 \AA, Eu {\sc ii}   4129 \AA, Ce {\sc ii}   4062 \AA,  Ce {\sc ii}   4083 \AA, Zr {\sc ii}  5350 \AA, etc.  (e.g. Manso Sainz et al. 2006).

Depolarizing collision  rates of complex ions are vital for a reliable interpretation of such lines and can be theoretically obtained only via our   semi-classical theory with good accuracy.  
 It is important to notice that it is very difficult or even impossible to treat collisional processes, involving complex
neutral and ionized atoms,   by standard quantum chemistry  methods. 
Our results allowed us to calculate, for the first time, the (de)polarization rates of the levels associated to complex atoms (Derouich et al. 2005a).  
\subsection{Accuracy of the  calculations}
 The results given by the  Equations (5--17) are concerned with simple atoms. The electronic configuration of the complex ion has one optical electron above an incomplete (i.e. open) subshell which has a non-zero angular momentum. To derive the collisional rates of levels of complex ions from those of simple ions, we consider that only the valence electron can undergo perturbations due to collisions with hydrogen atoms. This is  the frozen core approximation. This approximation has been adopted in our calculations of the depolarization and polarization transfer rates, but also it has been used by   ABO  semi-classical theory   for
collisional line broadening by neutral hydrogen.  

  ABO have tested the results for both complex atoms like Fe I (Multiplets 1146 and 1165) and simple atoms (Multiplets 7, 9 and 11 of Mg I; Multiplet 4 of Na I; Ca II infrared triplet). They    estimated the accurarcy of their theory to be around 20 \% or better (Barklem  1998).  
 
  ABO used two methods in  testing  the collisional rates. The rates were used to fit the spectrum of selected solar lines and the derived abundance  was compared with meteoritic abundance of the emitting atoms or the solar abundance obtained by other authors. Alternatively, they adopted  meteoritic abudance or previously determined solar abudance to obtain empirically collisional broadening cross-sections by fitting the line profile. In either method, a single parameter  is adjusted to achieve the best possible match between the observed solar spectrum and the synthesized spectrum (Barklem  1998).

 In the description of the complex or simple atom/ion, we adopt the same description used by ABO.   For example, calculations of (de)polarization rates and broadening rates associated with Fe I lines are based on the same interaction potential and the same dynamics of collisions. There are no additional conceptual approximations  to  extend the ABO theory to the calculations of the depolarization rates performed by DSB.
Thus, the conclusions of ABO concerning the accurarcy remain valid for the case of (de)polarization rates.  
This means that for complex   atoms/ions, the accurarcy of the (de)polarization rates must be around 20 \% or better.   

 The main idea which allowed the extension of the results obtained for simple atoms/ions to complex atoms/ions is the use of the frozen core approximation. 
 This allows treatment of   heavy and/or complex atoms/ions    like Fe I, Ti I,  NdII, EuII, CeII, ZrII, etc.., whose collisional rates cannot be presently calculated via   quantum chemistry methods. The spectral lines of such atoms/ions   show significant polarization peaks in many spectral lines (see the atlases by Gandorfer 2000, 2002, 2005).  
 
 In Section 2  of Derouich et al. (2005a),  the DSB model for complex atom/ion  was explained. According to the DSB model,  the  electronic coonfiguration of a complex atom/ion is composed of three parts which were fully explained by Derouich et al. (2005a). As a result,  it was possible to show how to  derive the depolarization and polarization transfer rates of complex ions from the rates associated to simple ions -- the (de)polarization rate of a complex atom can be written as a linear combination of the (de)polarization rates of simple atom.  Thus, the relationships given in this paper  are easily applicable to the case of complex ions.

\section{Conclusions}
 This paper  completes the previous work of Derouich (2017) which was devoted to the case of the polarization of the $p$-states of ions. 
 The scattering polarization profiles of the solar ions have been (and are still) widely investigated 
both from   theoretical  and observational points of view.  To gain  complete understanding of the scattering  polarization in the photosphere of the Sun, one should obtain collisional rates for ion-hydrogen interactions.  Calculations of the (de)polarization   collisional rates affecting the levels of the solar ions,  first require   the calculation of the  atomic wavefunctions and interaction  potentials.  Second, the time-dependent Schr\"odinger   equations  have to be solved  on the basis of  the  set of   eigenfunctions. This is a    complicated  problem which we show that  could be avoided by applying general relationships provided in this paper.   

 In fact, in this paper, we provide usefull relationships which allow the calculations of any  depoalrization and polarization transfer rates of the $d$-states of complex and simple ions. The hyperfine levels are also treated in this work.  The establishement of the relationships was possible, thanks to the general nature of the DSB approach, but also to the very powerful GP methods.

\section*{Acknowledgements}
This work was supported by the Deanship of Scientific Research (DSR), King Abdulaziz University, Jeddah, under grant No. (D-049-130-1437). The authors, therefore, gratefully acknowledge the DSR technical and financial support.

\nocite*{}
\bibliographystyle{pasa-mnras}

\bibliography{1r_lamboo_notes}

\end{document}